# Evolution of topological phases in atomically thin WTe$_2$ films

Changcang Qiao(乔昌仓)[1*], Chen-Chia Hsu(許宸嘉)[2*], Tao Zhang(张韬)[1], Zhiming Sun(孙志明)[1], Dong Qian(钱冬)[1], Yang-hao Chan(詹杨皓)[2#], Peng Chen(陈鹏)[1#]

[1]*Key Laboratory of Artificial Structures and Quantum Control (Ministry of Education), Tsung-Dao Lee Institute, School of Physics and Astronomy, Shanghai Jiao Tong University, Shanghai 200240, China.*

[2]*Institute of Atomic and Molecular Sciences, Academia Sinica, Taipei 10617, Taiwan.*

*These authors contributed equally to this work.

#Email: yanghao@as.sinica.edu.tw;

pchen229@sjtu.edu.cn

**Topological materials ranging from topological insulators to semimetals host many novel quantum phenomena including quantum spin Hall effect and topological Fermi arcs. Transitions between these topological phases have attracted much research interest. We performed angle-resolved photoemission spectroscopy (ARPES) on $WTe_2$ ranging from a monolayer to the bulk and reveal the evolution of the electronic structure and the band gap. Notably, the gap observed in the monolayer system is suppressed in the three layers, where the film becomes metallic. Variations in the topological properties with thickness are demonstrated by the first-principles calculations. Topological $Z_2$ invariant is shown to oscillate between 1 and 0 with the addition of layers, originating from the interlayer coupling-induced change in band crossing. The system evolves into a Weyl semimetal when the conduction and valence bands touch near the Fermi level and the topological nature is described by the Chern number. Our findings demonstrate the non-monotonic dependence of topological states on dimensionality and how layer-driven electronic band reconfiguration leads to phase transitions in solids.**

In materials, the non-trivial topology of the electronic bands characterized by topological invariants that remain unchanged under smooth transformations has attracted broad interest [1-6]. This field advanced with the discovery of topological insulators that are insulating in bulk and conducting at surfaces/edges [1-4]. In two-dimensional topological insulators, the metallic edge states are protected by time-reversal symmetry and remain robust against weak disorder, making them promising candidates for low-dissipation spintronic devices [7-10]. Further research revealed topological semimetals, where bands touch near the Fermi energy, including Dirac semimetals, Weyl semimetals, nodal-line semimetals, and systems with multi-fold band degeneracies [11-20]. The exotic electronic structures in topological semimetals lead to intriguing physics, such as negative magneto-resistance, topological Fermi arc surface states, chiral magnetic effects, and anomalous Hall effect [13, 21, 22]. These novel quantum phenomena are of fundamental interest for condensed matter physics and hold potential for advanced technological applications.

Transition metal chalcogenides contain rich elemental compositions and diverse structural phases, making them promising platforms for realizing various topological states. A notable example is the $MX_2$ family (M = Mo or W; X = S, Se, or Te) [23-32]. Among them, bulk $WTe_2$—a prototypical topological Weyl semimetal—has attracted much interest because of the giant magnetoresistance at low temperatures, which has been attributed to a compensation effect involving the electrons and holes in the system [27,28]. It crystallizes in an orthorhombic structure with space group $Pnm2_1$. The lack of inversion symmetry leads to the separation of the Weyl points of opposite chirality [33].

This layered material can be readily exfoliated into monolayer form or grown via molecular beam epitaxy (MBE). The monolayer $WTe_2$ is stable in 1T' phase with inversion symmetry. It is identified as a quantum spin Hall insulator up to 100 K with a sizable bulk gap [24,25]. In contrast, exfoliated bilayer $WTe_2$ exhibit broken inversion symmetry and function as ferroelectric insulators, demonstrating a finite polarization at low temperatures [34]. Exploring the layer-dependent electronic structure in these multilayer systems and how the topological nature evolves from a quantum spin Hall insulating state in the monolayer to a Weyl semimetal phase in the bulk is essential for understanding the mechanism of the topological phase transitions.

In this work, we fabricate few-layer $WTe_2$ films by MBE and systematically compare the band structure of $N$-layer films ($N$ = 1, 2, 3) with that of the bulk. First-principles calculations reveal the Td structure with broken inversion symmetry is energetically more favorable than 1T' structure in multilayer configurations. We observed a reduction of the bulk gap with increasing the thickness and the system evolves into a semimetal at $N = 3$. Topological $Z_2$ invariant is shown to oscillate between 1 and 0 with the addition of layers. A Weyl semimetal phase emerges when the conduction and valence bands touch around the Fermi level, marking the transition to a bulk-like electronic structure.

High-quality monolayer and multilayer $WTe_2$ films were successfully grown on bilayer-graphene-terminated 6H-SiC(0001) substrates using van der Waals epitaxy. Reflection high-energy electron diffraction (RHEED) revealed sharp diffraction patterns for a monolayer $WTe_2$ film grown at 250 °C [Fig. 1(b)], indicating a well-ordered crystalline structure. Core-level scans [Fig. 1(c)] exhibited characteristic Te $4d$ and W $4f$ peaks, with no peak splitting confirming the T' phase is the only phase at this growth temperature. Figure 1(e) shows the ARPES maps on monolayer $WTe_2$ taken along the $\overline{\Gamma X}$ direction. The overall band dispersion is consistent with the prior studies and theoretical calculations [23,25]. A hole-like valence band is observed centered around the $\overline{\Gamma}$ point. The faint conduction bands at -30 meV below the Fermi level is revealed in detailed high-resolution spectra [Fig. 2(a)]. The system is metallic because of the n-type self-doping during the growth. The indirect gap between the valence band maximum and the conduction band minimum is determined to be ~55 meV. In contrast, the ARPES spectra of bulk $WTe_2$ show a semimetallic character, with the valence bands cross over the Fermi level and an overlap between valence and conduction bands from the calculations.

To investigate how the band structure evolves from monolayer to bulk, layer-resolved ARPES measurements were performed on $WTe_2$ with thickness of $N = 1$-3 layers. The spectra taken with the unpolarized light from a He-discharge lamp at 10 K are shown in Fig. 2. For comparison, the first-principles calculations for freestanding multilayer $WTe_2$ ($N = 1$-3) are shown in Fig. 2(b). Density functional theory with two different functionals (GGA+U and HSE06) are carried out and the results agree well with the experiments (Fig. S7). Given the van der Waals nature of the interlayer coupling, the influence of the underlying graphene substrate is expected to be negligible. The band structure shows layer-by-layer evolution including the emergence of quantum well states as the film thickness increases. For instance, a quantum well peak appears at ~-0.3 eV in the bilayer sample ($N$ = 2). The valence band top in bilayer sample shifts closer to the Fermi level and the extracted band gap from line shape fitting to the EDCs is ~30 meV, which is smaller than the monolayer case [35]. The valence band crosses the Fermi level in the $N$ = 3 layers film and the gap is closed in consistent with the calculated results. These results indicate the $N$ = 3 layers system is already bulklike. The metallicity observed in our trilayer system is consistent with previous reports on exfoliated samples [34, 36]. Note that the conduction band in trilayer films is not clearly resolved due to the high background signal. Some of the subbands in $N$ layers predicted in the calculations exhibit weak intensity or remain unresolved in the ARPES spectra, which can be attributed to the matrix element effects. The ARPES spectra show a more parabolic shape compared to the relatively flat dispersion obtained from calculations. This renormalization is likely influenced by $n$-type self-doping during film growth. As demonstrated in the doping-dependent measurements (Supplemental Material, Fig. S3), increasing electron doping progressively sharpens the valence band top.

Upon warming to room temperature, the band structures of $N$ = 1 and 2 layers remain unchanged without noticeable variations (Supplemental Material, Fig. S1 and S2 [37]). Specifically, the band gap is stable up to 300 K in the ARPES measurements, suggesting the QSH effect in the monolayer system should be robust with temperature. However, the quantized conductance is only observed below 100 K in transport measurements [24], which can be attributed to the gap size of 55 meV—comparable to the thermal energy at room temperature. The absence of band renormalization with temperature demonstrates the excitonic effects are negligible in these *n*-type doped samples.

Furthermore, we performed Rb doping to shift the conduction bands downward the Fermi level (Supplemental Material, Fig. S3 [37]). The bulk gap of the monolayer sample decreases at a higher doping level and the conduction bands eventually overlap with the valence bands at a carrier density of $8.9\times10^{13}/cm^2$. Moreover, the band shapes become noticeably different such as the sharpening of the band top and reduced separation between the two highest valence bands. This doping effect cannot be simply described with a rigid shift of the bands. It may be originated from structural modifications caused by the intercalation of Rb in the interface between the film and the substrate. The ability to tune the QSH gap in the monolayer film with doping can be useful for controlling the QSH channels.

Topological property of monolayer $WTe_2$ is characterized by a band inversion between W 5*d* states with opposite parities near the zone center, which gives rise to a gap of 42 meV from the calculated PBE+U results (Fig. 2(b), Supplemental Material, Fig. S4 [37]). The nontrivial topology is further confirmed by the computed $Z_2$ invariant ($Z_2 = 1$), obtained via the evolution of hybrid Wannier charge centers, which exhibits a crossing between the middle of the largest gap and the hybrid Wannier charge center near the $\bar{\Gamma}$ point. This unequivocally identifies the monolayer as a quantum spin Hall (QSH) insulator. Consequently, topologically protected one-dimensional edge states are present within the bulk gap, as directly evidenced by the calculated edge spectrum in Fig. 3. In contrast, bilayer $WTe_2$ is found to be topologically trivial ($Z_2 = 0$). The edge state that connects the valence and conduction bands in the monolayer becomes two separate states. In contrast to the monolayer, these states do not bridge the gap but are instead entirely terminated within the valence or conduction bands, resulting in a trivial insulating phase.

This theoretical result is consistent with experimental measurements that showing no evidence of quantized edge conduction in bilayers [38].

The situation is different in trilayer WTe$_2$ as there is no global band gap in the band structure. As the conduction bands are not well resolved in the ARPES, the topological properties of trilayer films are discussed based mainly on the calculated results. Despite this metallic character, the valence and conduction bands remain separated in momentum space near the zone center. The well-separated bands allow for the well-defined calculation of the $Z_2$ invariant, and the result ($Z_2$ = 1) reveals the system's nontrivial topology despite its overall semimetallic character. The change of edge states with the number of layers implies that the $Z_2$ index is expected to oscillate between 1 and 0 as layers are added. This behavior ends in the bulk (3D) crystal of WTe$_2$, in which the conduction and valence bands cross and form type-II Weyl points [29,33]. The topological description thus shifts from the $Z_2$ invariant to the Chern number, which quantified the Berry flux around these band-touching points. The evolution of the topological state across the dimensional range from 2D limit to 3D is summarized in Fig. 4.

Our successful fabrication of multi-layer WTe$_2$ films combined with ARPES measurements enables direct observation of the electronic band structure across different layer thicknesses. First-principles calculations further reveal that the evolution of topological phases with increasing layer number is intimately related to the changes in the electronic bands. These results demonstrate that dimensionality serves as an effective tuning parameter for driving topological phase transitions in van der Waals materials. Other external parameters, such as strain or electric polarization, could also modify the band structure and potentially drive a topological phase transition [39-41].

**Acknowledgments**

We thank Prof. Hengxin Tan for useful discussion. The work at Shanghai Jiao Tong University is supported by the Ministry of Science and Technology of China under Grant No. 2022YFA1402400, and the National Natural Science Foundation of China (Grant No. 12374188). Y.H.C. acknowledges support from the Ministry of Science and Technology, the Academia Sinica (Project No. AS-CDA-114-M04), and the National Center for High-performance Computing in Taiwan.

**Fig. 1.** Film structure, and electronic band structure of WTe$_2$. (a) Views of the atomic structure of monolayer WTe$_2$. (b) RHEED patterns taken from a monolayer sample. (c) A core level scan taken with 100 eV photons. (d) Corresponding 2D Brillouin zones with high symmetry points labeled. ARPES maps taken along the $\overline{\Gamma X}$ direction from (e) the monolayer and (f) bulk samples at 10 K.

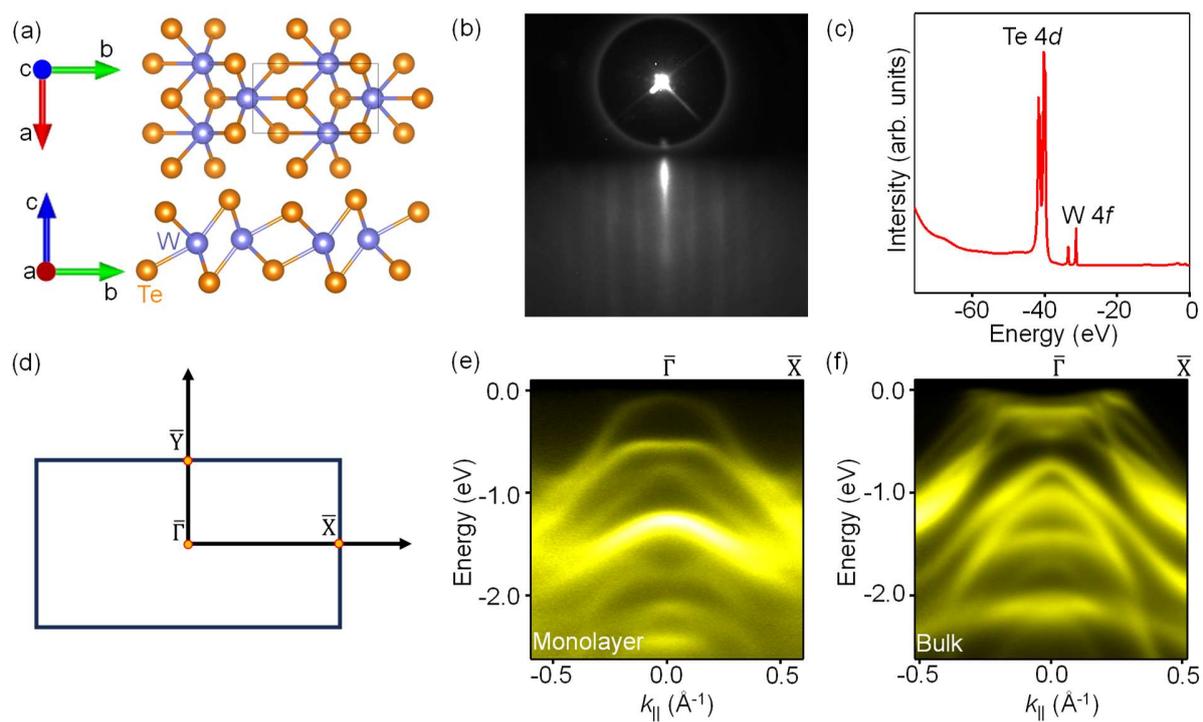

**Fig. 2.** Electronic band structure of WTe$_2$ thin films. (a) ARPES spectra of WTe$_2$ thin films with thickness ranging from 1 to 3 layers at 10 K. Calculated bands (red curves) with GGA + U method are overlaid on top of the spectra. (b) Corresponding second derivative spectra for comparison. (c) Calculated band structures with GGA+U method.

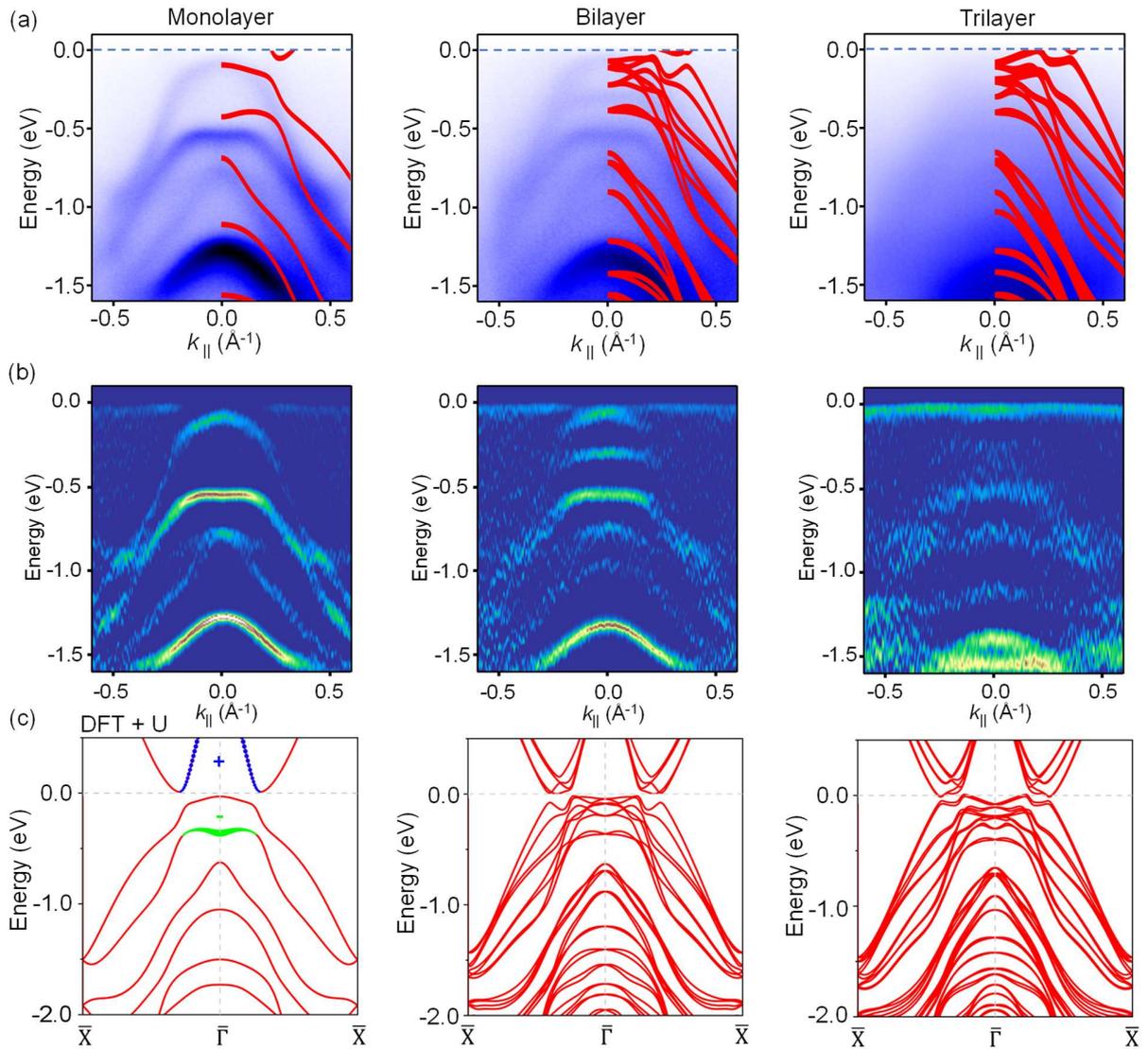

**Fig. 3.** Topological properties of ultrathin WTe$_2$ ($N$ = 1-3). (a)-(c), Calculated edge states along the $\overline{\Gamma X}$ direction. (d)-(f) Evolution of Wannier charge center along the $\overline{\Gamma X}$ direction. For $N$ = 1 and 3, the solid blue line crosses over the band odd times, indicating $Z_2$ = 1.

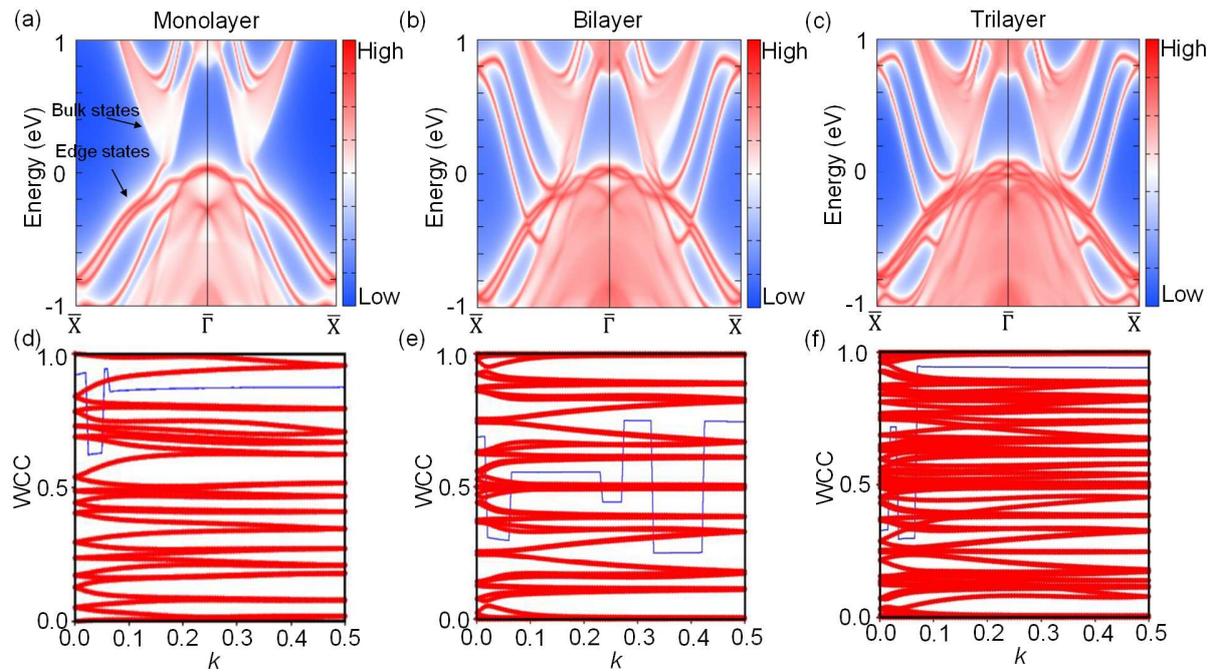

**Fig. 4.** Topological properties phase diagram of WTe$_2$. The red pentagrams denote the band gap as a function of thickness.

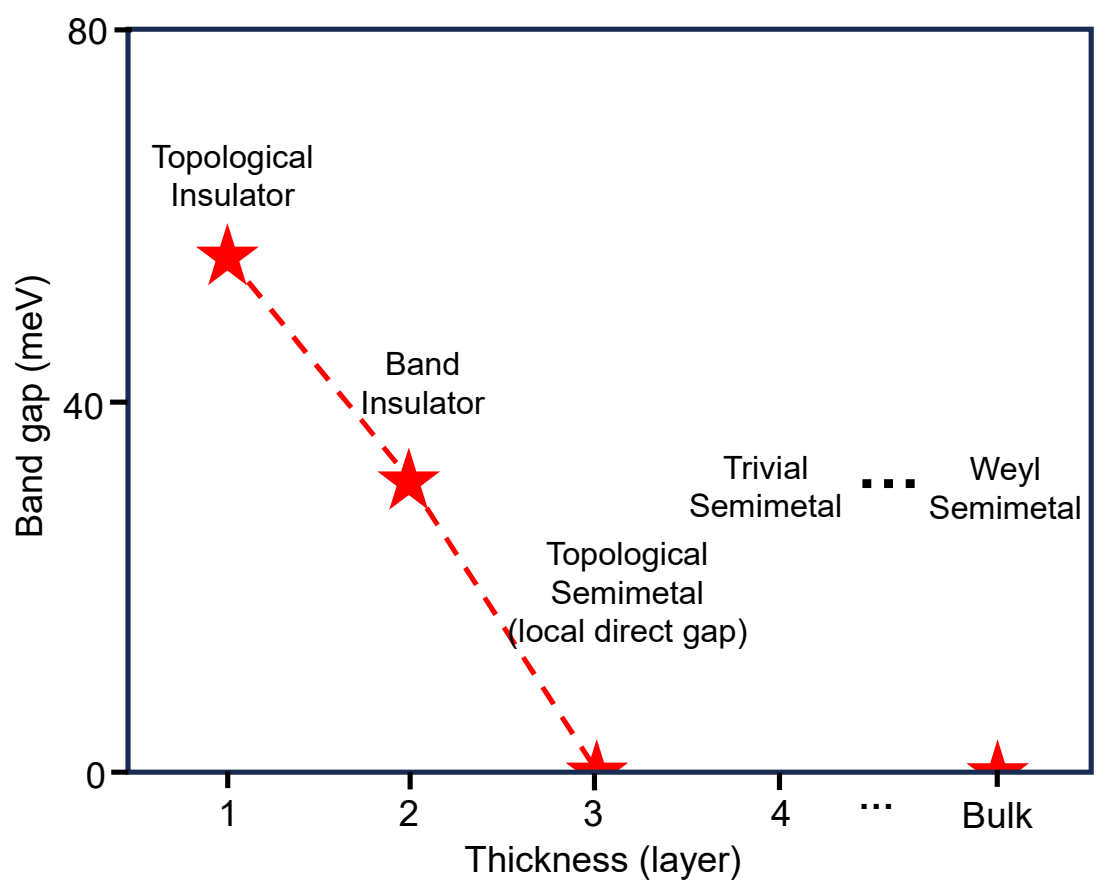